\documentclass[12pt]{article}

\usepackage{epsfig}
\usepackage{amsbsy}
\usepackage{amsmath}

\begin{document}

\title{On the determinations of the size and shape of the
interaction region from Bose-Einstein correlations}
\author{K.Zalewski
\\ M.Smoluchowski Institute of Physics
\\ Jagellonian University, Cracow\footnote{Address: Reymonta 4, 30 059 Krakow,
Poland, e-mail: zalewski@th.if.uj.edu.pl. This work has been partly supported
by the Polish Ministry of Education and Science grant 1P03B 045 29(2005-2008).
}
\\ and\\ Institute of Nuclear Physics, Cracow}
\maketitle

\begin{abstract}
Determinations of the size and shape of the interaction region from
$k$-particle ($k=1,2,\ldots$) momentum distributions of identical particles are
analyzed. The full group of transformations changing the single particle
density matrix without affecting any of the measurable momentum distributions
is identified. The corresponding uncertainties in the inferred parameters of
the interaction region are discussed.
\end{abstract}
\noindent PACS numbers 25.75.Gz, 13.65.+i \\Bose-Einstein correlations,
interaction region determination.

\section{Introduction}

The study of Bose-Einstein correlations, or more generally of single particle,
two particle etc. momentum distributions for identical particles, is the
standard tool used to learn about the size and shape of the interaction region
where the hadrons are produced. We will explicitly discuss spin zero bosons,
because in practice this is the most important case. Much of the argument,
however, can be extended to bosons with non-zero spins and to fermions.

Most published analyses are variants of the following approach which we will
also follow. One defines a single particle density matrix in the momentum
representation $\rho(\textbf{p};\textbf{p}')$. The diagonal elements of the
(symmetrized) $n$-particle density matrix are taken as \cite{KAR}

\begin{equation}\label{karczm}
\rho(\textbf{p}_1,\ldots , \textbf{p}_n;\textbf{p}_1,\ldots ,\textbf{p}_n) =
\sum_P\prod_{j=1}^n \rho(\textbf{p}_j;\textbf{p}_{Pj}),
\end{equation}
where the summation is over all the permutations $j \rightarrow Pj$ of the
indices $j$. The $n$-particle momentum distribution is given by the diagonal
elements of this $n$-particle density matrix\footnote{One can do somewhat
better by including more-particle density matrices, but these corrections,
known as multiparticle corrections,  are usually neglected, because they are
not very important (cf. e.g. the review \cite{WIH}).}. Thus, all the
information about the system is contained in the single-particle density matrix
$\rho(\textbf{p};\textbf{p}')$. The question is: how well this function can be
determined from the experimental data?

The single particle momentum distribution is

\begin{equation}\label{}
  P(\textbf{p}) = \rho(\textbf{p};\textbf{p}).
\end{equation}
Thus the diagonal elements are measurable. The two-particle distribution is

\begin{equation}\label{}
  P(\textbf{p}_1,\textbf{p}_2) = \rho(\textbf{p}_1;\textbf{p}_1)\rho(\textbf{p}_2;\textbf{p}_2) +
  |\rho(\textbf{p}_1;\textbf{p}_2)|^2,
\end{equation}
where the hermiticity of the density matrix

\begin{equation}\label{rhoher}
  \rho(\textbf{p}_1;\textbf{p}_2) = \rho^*(\textbf{p}_2;\textbf{p}_1)
\end{equation}
has been used. Thus, the absolute value of every matrix element is also
measurable. For the three-particle momentu distribution formula (\ref{karczm})
yields

\begin{equation}\label{}
  P(\textbf{p}_1,\textbf{p}_2,\textbf{p}_3) = \ldots +
  2\Re\left[\rho(\textbf{p}_1;\textbf{p}_2)\rho(\textbf{p}_2;\textbf{p}_3)\rho(\textbf{p}_3;\textbf{p}_1)\right],
\end{equation}
where the dots denote terms which can be determined from the single particle
and two-particle momentum distributions and $\Re $ stands for real part of. In
general, since every permutation can be decomposed into cycles, the
$n$-particle momentum distribution is

\begin{equation}\label{}
  P(\textbf{p}_1,\ldots,\textbf{p}_n) = \ldots + 2\Re\left[\rho(\textbf{p}_1;\textbf{p}_2)\rho(\textbf{p}_2;
  \textbf{p}_3)\ldots\rho(\textbf{p}_n;\textbf{p}_1)\right],
\end{equation}
where the dots denote terms which can be determined by measuring momentum
distributions for less than $n$ particle.

In a previous paper \cite{BIZ} it has been pointed out that the measurable
quantities do not change when the substitution

\begin{equation}\label{}
  \rho(\textbf{p};\textbf{p}') \rightarrow e^{i\left(f(\textbf{p}_1) - f(\textbf{p}_2)\right)}\rho(\textbf{p};\textbf{p}')
\end{equation}
is made\footnote{This remains valid also when the multiparticle corrections are
included.}. In the present paper, we show that this group of transformations
supplemented by the transformation

\begin{equation}\label{}
\rho(\textbf{p};\textbf{p}') \rightarrow \rho^*(\textbf{p};\textbf{p'}) =
\rho(\textbf{p}';\textbf{p})
\end{equation}
generates the full group of transformations which leave the measurable
distributions invariant.

We will discuss the following problem. Somebody found that a density matrix
$\rho_0(\textbf{p};\textbf{p}')$ gives the best fit to all the measured
momentum distributions. What other density matrices
$\rho(\textbf{p};\textbf{p}')$ give exactly the same fit and how the
conclusions concerning the interaction region are affected when
$\rho_0(\textbf{p};\textbf{p}')$ is replaced by $\rho(\textbf{p};\textbf{p}')$?
This is the problem of the theoretical uncertainty in the determination of the
parameters of the interaction region from the measured momentum distributions.

\section{What can and what cannot be measured}

Let us write the single particle density matrix in the form

\begin{equation}\label{}
  \rho(\textbf{p}_1;\textbf{p}_2) = |\rho(\textbf{p}_1;\textbf{p}_2)|e^{i\chi(\textbf{p}_1,\textbf{p}_2)}.
\end{equation}
The modulus is measurable, thus all the ambiguities result from the phase. It
is convenient to choose for the phase the convention

\begin{equation}\label{}
  -\pi \leq \chi(\textbf{p},\textbf{p}') \leq \pi.
\end{equation}
The hermiticity of the single particle density matrix implies that

\begin{equation}\label{chiher}
  \chi(\textbf{p},\textbf{p}') = -\chi(\textbf{p}',\textbf{p}).
\end{equation}
Moreover, the phases $\chi(\textbf{p}_j,\textbf{p}_k)$ are constrained by the
measurable three-particle momentum distribution which yields

\begin{equation}\label{phase3}
  \cos\left[\chi(\textbf{p}_1;\textbf{p}_2) +
  \chi(\textbf{p}_2;\textbf{p}_3)+\chi(\textbf{p}_3,\textbf{p}_1)\right].
\end{equation}
The sign of the argument of the cosine is not constrained. This implies that
$\rho(\textbf{p};\textbf{p}')$ can be replaced by its complex conjugate without
affecting the fit to the data.

Once the sign is fixed, it is possible to determine from the data the function

\begin{equation}\label{inhequ}
  \Psi(\textbf{p}_1,\textbf{p}_2,\textbf{p}_3) = \chi(\textbf{p}_1;\textbf{p}_2) +
  \chi(\textbf{p}_2;\textbf{p}_3)+\chi(\textbf{p}_3,\textbf{p}_1).
\end{equation}
Formally, this is an inhomogeneous, linear functional equation for the function
$\chi(\textbf{p},\textbf{p}')$. Let us assume that some function
$\chi_0(\textbf{p},\textbf{p}')$ is a solution of this equation. If the model
agrees with experiment such a function must exist. Let us define further a
function $\chi_{h}(\textbf{p},\textbf{p}')$ which is a solution of the
homogenous equation

\begin{equation}\label{phahom}
\chi_{h}(\textbf{p}_1;\textbf{p}_2) +
  \chi_h(\textbf{p}_2;\textbf{p}_3)+\chi_h(\textbf{p}_3,\textbf{p}_1) = 0.
\end{equation}
Function $\chi_0(\textbf{p},\textbf{p}') + \chi_h(\textbf{p},\textbf{p}')$ is
then also a solution of equation (\ref{inhequ}). Thus, the question what is the
uncertainty of solution $\chi_0$ reduces to the question what is the full set
of solutions of equation (\ref{phahom}).

Differentiating both sides of this equation with respect to $p_{1j}$ and
$p_{2k}$ we get the equations

\begin{equation}\label{equdif}
  \frac{\partial \chi_h(\textbf{p},\textbf{p}')}{\partial p_j \partial p'_k} = 0,\qquad j,k = x,y,z.
\end{equation}
The general solution of this equation system can be written in the form

\begin{equation}\label{soldif}
  \chi(\textbf{p},\textbf{p}') = f(\textbf{p}) - g(\textbf{p}'),
\end{equation}
where $f(\textbf{p})$ and $g(\textbf{p})$ are arbitrary functions. Every
solution of equations (\ref{phahom}) is a solution of equation (\ref{equdif}),
but not the other way round. Substituting solution (\ref{soldif}) into equation
(\ref{phahom}) one easily finds that the equation is satisfied if and only if

\begin{equation}\label{}
  g(\textbf{p}) = f(\textbf{p}).
\end{equation}
The same conclusion follows from the symmetry (\ref{chiher}).  Thus, the most
general single particle density matrix giving the same fit to the data as
matrix $\rho_0(\textbf{p},\textbf{p}')$ can be written in the form

\begin{equation}\label{genden}
  \rho(\textbf{p};\textbf{p}') = e^{i\left(f(\textbf{p}_1) -
  f(\textbf{p}_2)\right)}\rho_0(\textbf{p};\textbf{p}')\quad \mbox{or}\quad
   e^{-i\left(f(\textbf{p}_1) -
   f(\textbf{p}_2)\right)}\rho^*_0(\textbf{p};\textbf{p}'),
\end{equation}
where $f(\textbf{p})$ is an arbitrary real function of the momentum
$\textbf{p}$. The four- and more particle distributions do not introduce any
more constraints.

To see how this ambiguity can be eliminated by a theoretical assumption let us
suppose that there are no position-momentum correlations. Then the Wigner
function (cf. (\ref{wigfun})) is a product of a function of $\textbf{x}$ only
and a function of $\textbf{K}$ only. Consequently, each of the density matrices
$\rho(\textbf{p};\textbf{p}')$ and $\rho_0(\textbf{p};\textbf{p}')$ is a
product of a function of $\textbf{p} + \textbf{p}'$ only and a function of
$\textbf{p} - \textbf{p}'$ only. This implies that

\begin{equation}\label{}
  f(\textbf{p}) - f(\textbf{p}') = \textbf{a}\cdot (\textbf{p} - \textbf{p}'),
\end{equation}
where $\textbf{a}$ is an arbitrary constant vector. The ambiguities in the
interaction region, corresponding to the transformations (\ref{genden}), reduce
to rigid translations by the vector $\textbf{a}$ and/or space inversion. Since
these do not affect the size and shape of the interaction region, one may claim
that there is no significant ambiguity left.

We conclude that, if somebody fitting the experimental data and using some
model (this is necessary to get a well-defined result), finds the effective
single particle density matrix $\rho_0(\textbf{p};\textbf{p}')$, another model
leading to any of the density matrices (\ref{genden}) agrees with the data just
as well. Of course, a priori some model may be more plausible than another, but
these are arguments on a different level -- unrelated to the fitting of
momentum distributions.

The effects of this uncertainty on the interpretation of the data can be
spectacular. It is plausible, for instance, that the momenta and positions at
birth of the hadrons are correlated so that \cite{CSZ}, \cite{BIZ1}

\begin{equation}\label{}
  \textbf{p} \approx \lambda \textbf{x},
\end{equation}
where $\lambda$ is a positive constant. Using the relation between the density
matrix and the Wigner function one easily finds that the complex conjugation of
the density matrix corresponds to the change of sign of the coordinate in the
Wigner function. Using further the interpretation of the Wigner function as an
approximation to the phase space distribution one finds that models where

\begin{equation}\label{}
 \textbf{p} \approx -\lambda \textbf{x}
\end{equation}
fit the data exactly as well as the previous ones.

\section{Moments of the particle density distribution in the interaction
region}

The density of particles in the interaction region is given by the diagonal
elements of the effective single particle density matrix in the coordinate
representation:

\begin{equation}\label{}
  \tilde{\rho}(\textbf{x};\textbf{x}) \equiv \tilde{\rho}(\textbf{x}) =
  \int\!\!\frac{dKdq}{(2\pi)^3}\;e^{i\textbf{qx}}\rho(\textbf{K},\textbf{q}),
\end{equation}
where

\begin{equation}\label{}
  \textbf{K} = \frac{1}{2}(\textbf{p} + \textbf{p}'),\qquad \textbf{q} = \textbf{p}-\textbf{p}'
\end{equation}
and $\rho(\textbf{K},\textbf{q})$ stands for  $\rho(\textbf{p};\textbf{p}')$.
Integrating $n$ times by parts over the components of $\textbf{q}$ and assuming
that $\rho(\textbf{K},\textbf{q})$ and all its derivatives with respect to
components of  $\textbf{q}$ tend to zero for $|\textbf{q}| \rightarrow \infty$
one finds in particular

\begin{equation}\label{}
  \tilde{\rho}(\textbf{x}) =
  \frac{i^n}{x_j^n}\int\!\!\frac{dKdq}{(2\pi)^3}\;e^{i\textbf{qx}}\left(\frac{\partial^n}
  {\partial q_j^n}\rho(\textbf{K},\textbf{q})\right), \quad j=x,y,z.
\end{equation}
Multiplying both sides of this identity by $x_j^n$ and integrating over
$\textbf{x}$ one gets the moments

\begin{equation}\label{}
  \int\!\!dx\;\tilde{\rho}(\textbf{x})x_j^n \equiv \langle x_j^n \rangle =
  i^n\int\!\!dK
  \left( \frac{\partial^n}{\partial q_j^n}\rho(\textbf{K},\textbf{q})\right)_{\textbf{q} = 0}.
\end{equation}
Usually only the first and second order moments are considered.

Let us introduce the following notation for the moments corresponding to the
density matrix $\rho_0(\textbf{p};\textbf{p}')$. Putting

\begin{eqnarray}\label{}
  r_{0j} &=& i\int\!\!dK
  \left( \frac{\partial}{\partial q_j}\rho_0(\textbf{K},\textbf{q})\right)_{\textbf{q} =
  0},\qquad j = x,y,z;\\
  R^2_{0jk} &=& -\int\!\!dK
  \left( \frac{\partial^2}{\partial q_j\partial q_k}\rho_0(\textbf{K},\textbf{q})\right)_{\textbf{q} =
  0}\qquad j,k = x,y,z
\end{eqnarray}
and assuming that the effective single particle density matrix
$\rho_0(\textbf{p}_1,;\textbf{p}_2)$ is the correct one, the moments are

\begin{eqnarray}\label{}
\langle \textbf{x} \rangle &=& \textbf{r}_0 \\
\langle x_jx_k \rangle &=& R^2_{0jk}.
\end{eqnarray}
It has been known for a long time (cf. e.g. \cite{WIH}) that it is not possible
to determine either $\langle \textbf{x} \rangle$ or $\langle x_jx_k \rangle$
from momentum measurements. The usual way out (cf. e.g. \cite{WIH}) is to
calculate instead the variances

\begin{equation}\label{}
  \sigma^2_0(x_j) = \langle x_j^2 \rangle - \langle x_j \rangle^2.
\end{equation}

Another set of much discussed quantities is related to the integrands of the
integrals given above

\begin{eqnarray}\label{}
  r_{0j}(\textbf{K}) &=& i
  \frac{\left( \frac{\partial}{\partial
  q_j}\rho_0(\textbf{K},\textbf{q})\right)_{\textbf{q}=0}}{\rho_0(\textbf{K},\textbf{0})}, \\
  R_{0jk}^2(\textbf{K}) &=& -\frac{\left( \frac{\partial^2}{\partial q_j\partial q_k}
  \rho_0(\textbf{K},\textbf{q})\right)_{\textbf{q} =0}}{\rho_0(\textbf{K},\textbf{0})}.
\end{eqnarray}
The corresponding variances and correlation functions are
\begin{equation}\label{}
 R^2_{0jk}(\textbf{K}) -
r_{0j}(\textbf{K})r_{0k}(\textbf{K}) = -\left(\frac{\partial^2}{\partial
q_j\partial q_k}\log \rho_0(K,q)\right)_{\textbf{q}=0}.
\end{equation}
Denoting the left-hand side of this equality by $R^2_{HBT,jk}$, and noting that
according to (\ref{chiher}) all the second derivatives of the phase with
respect to components of $q$ vanish at $\textbf{q} = \textbf{0}$, we get

\begin{equation}\label{}
R_{HBT,jk}^2(\textbf{K}) =  -\left(\frac{\partial^2}{\partial q_j\partial
q_k}\log |\rho_0(K,q)|\right)_{\textbf{q}=0}.
\end{equation}
It is seen that these functions are measurable and do not depend on the
specific choice of the function $\rho_0(\textbf{K},\textbf{x})$  among the
functions giving the same fits. They are just the well-known $HBT$ radii. We
have kept the traditional notation with the squares (cf. e.g. \cite{WIH}) in
spite of the well-known fact that $R^2_{HBT}$ for $j \neq k$ may be negative
(cf. e.g. \cite{WIH}). An important point is that, they are not sufficient to
calculate the overall variances and correlation functions because in general

\begin{equation}\label{}
  \langle r_{0j}(\textbf{K})r_{0k}(\textbf{K})\rangle \neq  \langle r_{0j}(\textbf{K})\rangle
  \langle r_{0k}(\textbf{K})\rangle.
\end{equation}

The equations for the moments become more suggestive when written in terms of
the Wigner function, related to the density matrix
$\rho_0(\textbf{K},\textbf{q})$ by

\begin{equation}\label{wigfun}
  \rho_0(\textbf{K},\textbf{q}) = \int\!\!dx \;W_0(\textbf{K},\textbf{x})e^{-i\textbf{qx}}.
\end{equation}
Then
\begin{eqnarray}\label{}
r_{0j}(\textbf{K}) &=& \frac{\int\!\!dx \;x_jW_0(\textbf{K},\textbf{x})}
{\int\!\!dx \;W_0(\textbf{K},\textbf{x})},\\
R^2_{0jk}(\textbf{K}) &=& \frac{\int\!\!dx
\;x_jx_kW_0(\textbf{K},\textbf{x})}{\int\!\!dx \;W_0(\textbf{K},\textbf{x})}.
\end{eqnarray}
Interpreting the Wigner function as the phase space density, as often done, one
sees that $r_{0j}(\textbf{K})$ and $R^2_{0jk}(\textbf{K})$ are the moments of
the components of $\textbf{x}$ evaluated at fixed $\textbf{K}$. This point of
views is supported by the fact that when averaged over $\textbf{K}$, i.e.
multiplied by the distribution of $\textbf{K}$ ($\rho_0(\textbf{K}) \equiv
\rho_0(\textbf{K},\textbf{0})$) and integrated over $\textbf{K}$, they
reproduce correctly the overall moments. However, according to Heisenberg's
uncertainty principle, when the variance of $\textbf{K}$ tends to zero the
variance of $\textbf{x}$ tends to infinity. Only when the variance of
$\textbf{K}$ grows from zero to finite values, destructive interference between
the contributions corresponding to different $\textbf{K}$ values limits the
$\textbf{x}$-range. Thus, the interpretation of $r_{0j}(\textbf{K})$ and
$R^2_{0jk}(\textbf{K})$ as moments of $\textbf{x}$ at given $\textbf{K}$ has
heuristic value, but should be used with care. Note also that the average over
$\textbf{K}$ of the variance $\sigma^2_K(x_j)$ does not give $\sigma^2(x_j)$.

In the following section we will discuss how the moments introduced here are
affected by the uncertainty in the phase of the density matrix
$\rho(\textbf{p},\textbf{p}')$.

\section{Uncertainties in the determination of the moments}

Let us see how the first and second order moments of the coordinates change
under transformations (\ref{genden}). The terms higher order than second in the
components of $\textbf{q}$ do not contribute and, therefore, the factors
multiplying $\rho_0(\textbf{K},\textbf{q})$ and
$\rho_0^*(\textbf{K},\textbf{q})$ can be written respectively as

\begin{equation}\label{}
  F_\pm(\textbf{K},\textbf{q}) = 1 \pm i\sum_j q_j\frac{\partial f(\textbf{K})}{\partial K_j}
  - \frac{1}{2}\sum_{j,k}q_jq_k\frac{\partial f(\textbf{K})}{\partial K_j}
  \frac{\partial f(\textbf{K})}{\partial K_k}.
\end{equation}

Using the formulae from the preceding section one finds

\begin{eqnarray}\label{}
r_j(\textbf{K}) &=& \pm \left( r_{0j}(\textbf{K}) - \frac{\partial f(\textbf{K})}{\partial K_j}\right),\\
R^2_{ij}(\textbf{K}) &=& R^2_{0jk}(\textbf{K})
-\left(r_{0j}(\textbf{K})\frac{\partial f(\textbf{K})}{\partial K_k} +
r_{0k}(\textbf{K})\frac{\partial f(\textbf{K})}{\partial K_j}\right) +\nonumber
\\& & \frac{\partial f(\textbf{K})}{\partial K_j}\frac{\partial
f(\textbf{K})}{\partial K_k} .
\end{eqnarray}
The \textit{HBT} radii remain, of course, unchanged.

The formulae for the overall moments are

\begin{eqnarray}\label{}
r_j &=& r_{0j} - \left\langle \frac{\partial f(\textbf{K})}{\partial K_j}
\right\rangle_K,\\
R^2_{jk} &=& R^2_{0jk} - \left\langle r_{0j}(\textbf{K})\frac{\partial
f(\textbf{K})}{\partial K_k} +
r_{0k}\frac{\partial f(\textbf{K})}{\partial K_j} \right\rangle_K + \nonumber \\
&& \left\langle \frac{\partial f(\textbf{K})}{\partial K_j}\frac{\partial
f(\textbf{K})}{\partial K_k} \right\rangle_K,
\end{eqnarray}
where $\left\langle \ldots \right\rangle_K$ denotes averaging over
$\textbf{K}$. The overall variances and correlation functions are

\begin{eqnarray}\label{}
  R_{jk}^2 - r_jr_k &=& \left\langle R^2_{HBT,jk}(\textbf{K}) \right\rangle_K +
  \left\langle \frac{\partial f(\textbf{K})}{\partial K_j}\frac{\partial f(\textbf{K})}{\partial
  K_k}\right\rangle_K
  - \left\langle \frac{ \partial f(\textbf{K})}{\partial K_j}  \right\rangle_K \left\langle \frac{
  \partial f(\textbf{K})}{\partial K_k}
    \right\rangle_K \nonumber \\
   && -
    \left\langle r_{0j}(\textbf{K})\frac{\partial f(\textbf{K})}{\partial K_k}\right\rangle_K
    + r_{0j}\left\langle\frac{\partial f(\textbf{K})}{\partial
    K_k}\right\rangle_K\nonumber \\ && -
    \left\langle r_{0k}(\textbf{K})\frac{\partial f(\textbf{K})}{\partial K_j}\right\rangle_K
    + r_{0k}\left\langle\frac{\partial f(\textbf{K})}{\partial K_j}\right\rangle_K.
\end{eqnarray}
They are not measurable without further assumptions.

Let us discuss some illustrative examples. For the models where

\begin{equation}\label{ridzer}
  r_0(\textbf{K}) \equiv \textbf{0},
\end{equation}
one gets

\begin{equation}\label{}
  R^2_{jj} = R^2_{HBT,jj} + \sigma^2_f(K_j), \qquad j=x,y,z,
\end{equation}
where the variance

\begin{equation}\label{}
\sigma^2_f(K_j) \equiv
 \left\langle \left(
 \frac{\partial f(\textbf{K})}{\partial K_j}
 \right)^2
  \right\rangle_K
 - \left\langle \frac{\partial f(\textbf{K})}{\partial K_j}\right\rangle_K\left\langle\frac{\partial
f(\textbf{K})}{\partial K_j} \right\rangle_K.
\end{equation}
It is seen that the \textit{HBT} radii are a rigorous lower bounds for the true
radii, while there is no upper bound. The interaction region can be assumed to
be the size of a football and agreement with experiment remains as good as
before. In practice, of course, so large interactions regions are excluded by
what is known about the detector, but this is information going beyond the
measurements of momentum distributions.

If a model giving exactly the same fit as the models satisfying (\ref{ridzer})
exists, it must be possible to find a function $f(\textbf{K})$ satisfying

\begin{equation}\label{}
  \nabla f(\textbf{K}) = \textbf{r}_0(\textbf{K}).
\end{equation}
Unless there are singularities, this equation can be solved when the vector
field $\textbf{r}_0(\textbf{K})$ is rotationless:

\begin{equation}\label{}
  \nabla \times \textbf{r}_0(K) = \textbf{0}.
\end{equation}

Consider the following model discussed by Akkelin and Sinyukov \cite{AKS}. A
classical gas of noninteractring particles has at time $t=0$ a Gaussian
distribution in coordinate space with $\langle \textbf{x} \rangle = 0$ and a
Boltzmann distribution in momentum space; there are no position-momentum
correlations. It is obvious how this gas will evolve. The momentum distribution
will not change. In coordinate space each group of particle having identical
momenta $\textbf{K}$ will propagate rigidly with velocity
$\frac{\textbf{K}}{m}$, where $m$ is the particle mass. This picture can be
confirmed by solving the corresponding Boltzmann equation \cite{AKS}. For the
particles with momentum $\textbf{K}$

\begin{equation}\label{}
  \langle \textbf{x} \rangle(\textbf{K}) = \frac{\textbf{K}}{m}t =
  \textbf{r}_0(\textbf{K}),
\end{equation}
where the second equality corresponds to the choice $f(\textbf{K}) = 0$, and
corresponds to

\begin{equation}\label{}
  R^2_{jk} = R^2_{HBT,jk}.
\end{equation}

We can choose, however, just as well

\begin{equation}\label{}
  f(\textbf{K}) = \frac{\textbf{K}^2}{2m}t.
\end{equation}
The corresponding density matrix satisfies (\ref{ridzer}) and consequently
yields

\begin{equation}\label{}
  R^2_{jj} = R^2_{HBT,jj} + \left\langle \frac{K^2_j}{m^2}t^2 \right\rangle_K.
\end{equation}
From the description of the model, which contains more information than just
about the momentum distributions, we know that the first description is wrong
and the second is correct. In this example the \textit{HBT} radii can be
interpreted as follows. For particles with any given value of $\textbf{K}$
there is some distribution of $\textbf{x}$. Shifting rigidly all these
$\textbf{x}$ distributions on top of each other so that their averages $\langle
\textbf{x} \rangle(K)$ coincide one gets a reduced distribution. The
\textit{HBT} radii are the radii of this reduced distribution.

In \cite{BIZ1} a model was proposed, where for the transverse degrees of
freedom

\begin{equation}\label{}
  \langle x_j \rangle(K_j) = \lambda K_j
\end{equation}
and $\lambda$ does not depend on the transverse momenta. Again, using the
description of the model one finds

\begin{equation}\label{}
  f(K_j) = \frac{\lambda}{2}K_j^2
\end{equation}
and

\begin{equation}\label{}
  R^2_{jj} = R^2_{HBT,jj} + \lambda^2\left\langle K_j^2\right\rangle_K.
\end{equation}
This was used to explain why experimentally the $HBT$ radii decrease when the
particle mass increases. In this model the true radii $R^2_{jj}$ do not depend
on the particle mass and the observed decrease of the $HBT$ radii with
increasing particle mass just reflects the well-known mass dependence of the
second term on the right-hand side.

\section{Conclusions}

We have considered the most popular class of models, where all the information
about the size and shape of the interaction region is contained in a
single-particle density matrix $\rho(\textbf{p};\textbf{p}')$. The aim is to
determine this matrix from momentum distributions measured for sets of
identical particles. We show that, without further assumptions, this is not
possible. For each single particle density matrix there is an equivalence class
of matrices which give exactly the same fits to all the momentum distributions.
All these matrices are related by the transformations (\ref{genden}). Thus, the
full group of transformations leaving the momentum distributions invariant has
been identified.

The $HBT$ radii, which can be unambiguously determined from the momentum
distributions, are in general different from the true radii of the interaction
region. Sometimes, but not always, they are lower bounds for the true radii. In
order to determine the true radii it is necessary to determine a function
$\nabla f(\textbf{K})$. If only the results of momentum measurements are
available function $f(\textbf{K})$ is completely arbitrary. Models, however,
usually contain information going beyond that obtainable from momentum
measurements and then function $\nabla f(\textbf{K})$ may be possible to
determine. It is recommended, when proposing a model, to formulate clearly and
justify the assumptions which yield $f(\textbf{K}) = 0$, because these
assumptions cannot be justified even by the best fits to the measured momentum
distributions.

We have discussed a comparatively simple version of the standard approach. It
is easy to extend our analysis to include the multiparticle corrections. The
final state interactions and, more generally, interparticle correlations,
however, would make the discussion much more complicated. It is most unlikely,
nevertheless, that they would make the result simpler. Ambiguities analogous to
those described here are expected to occur also in these more complicated
models.


\end{document}